\documentclass{elsart}

\usepackage{graphicx}
\usepackage{amssymb}

\begin{document}

\begin{frontmatter}

\title{Anticipating the dynamics of chaotic maps}

\author[E]{Emilio Hern\'{a}ndez-Garc\'{\i}a},
\ead{emilio@imedea.uib.es}
\author[M,C]{C. Masoller},
\ead{cris@fisica.edu.uy}
\author[C]{Claudio R. Mirasso}
\ead{claudio@imedea.uib.es}

\address[E]{Instituto Mediterr\'{a}neo de Estudios Avanzados, IMEDEA
(CSIC-UIB), Campus Universitat de les Illes Balears, E-07071 Palma
de Mallorca, Spain}

\address[M]{Instituto de F\'{\i}sica, Facultad de Ciencias,
Universidad de la Rep\'ublica, Igu\'a  4225, Montevideo 11400, Uruguay}

\address[C]{Departament de F\'{\i}sica, Universitat de les
Illes Balears, E-07071 Palma de Mallorca, Spain}

\begin{abstract}
We study the regime of anticipated synchronization in
unidirectionally coupled chaotic maps such that the slave map has
its own output reinjected after a certain delay. For a class of
simple maps, we give analytic conditions for the stability of the
synchronized solution, and present results of numerical
simulations of coupled 1D Bernoulli-like maps and 2D Baker maps,
that agree well with the analytic predictions.
\end{abstract}

\begin{keyword}
Chaos synchronization \sep Anticipated synchronization

\PACS 05.45.Xt \sep 05.45.Gg

\end{keyword}

\date{26 October 2001}
\end{frontmatter}

The synchronization of chaotic systems is a subject that has
attracted a lot of attention in the past years. Since the
pioneering works [1] several different regimes of synchronization
have been found: complete synchronization, phase synchronization
[2], lag synchronization [3], generalized synchronization [4,5],
synchronization by common noise force [6,7], among others.

Anticipated synchronization is a recently discovered
synchronization regime that occurs in unidirectionally coupled
systems [8,9]. In this regime counterintuitive phenomena occur,
since the slave system anticipates the chaotic evolution of the
master system, despite the fact that chaotic behavior implies
long-term unpredictability.

In the case of coupled time-delayed differential equations the
anticipation time is related to the difference between the
intrinsic delay time of the systems and the delay time of the
coupling [10]. In coupled ordinary differential equations (as the
Lorenz and Rossler systems) the anticipation time must be small to
have a stable synchronization manifold [8]. However, by using a
chain of slave systems, anticipation times that are multiples of
the coupling delay time and that exceed characteristic time scales
of the chaotic dynamics can be obtained [9,11]. Many additional
numerical [12,13] and experimental [14,15] studies of anticipated
synchronization have been performed.

Recently, analytic conditions for the synchronization of coupled
maps with delays were given by Masoller and Zanette [16]. In the
maps considered in [16] the chaotic behavior is induced by the
delay term in the map (in other words, without delay the master
map is not chaotic). In Ref. [16] the master map ($x_n$) and the
slave map ($y_n$) are of the form
\begin{equation}
x_{n+1}=bx_n + f(x_{n-N}) ,
\end{equation}
\begin{equation}
y_{n+1}=by_n + (1-\eta)f(y_{n-N}) + \eta f(x_{n-M}) .
\end{equation}
with $|b|<1$ and $f$ a nonlinear function. The synchronized
solution is given by $y_n=x_{n-M+N}$. By studying the evolution of
a small perturbation of the solution it was shown that the
synchronization is stable for $\eta=1$. The existence of a
threshold value of the coupling, $\eta_c<1$, above which the
synchronized solution is stable, was also shown.

When the chaotic behavior of the master map is not induced by a
delay, to the best of our knowledge no analytical conditions for
anticipation have been reported. In this letter we study a system
composed by a chaotic map (master) unidirectionally coupled to a
second chaotic map (slave) which has its own signal reinjected
after a certain delay. We consider a master map, $x_n$, and a
slave map, $y_n$ of the form
\begin{equation}
x_{n+1}= f(x_n) ,
\end{equation}
\begin{equation}
y_{n+1}= f(y_n) + \gamma(x_{n-N} - y_{n-M}) ,
\end{equation}
where $f$ is a nonlinear function. We show that the delayed
auto-injection in the slave map leads to anticipation in the
synchronization, and present analytic conditions for the stability
of anticipated synchronization. An analytical treatment is
possible because we consider simple chaotic maps. We exemplify the
results with numerical simulations of coupled 1D Bernoulli-like
maps and 2D Baker maps.

The synchronization manifold of (3), (4) is given by
\begin{equation}
y_{n}= x_{n-N+M} ,
\end{equation}
and thus the slave variable is lagged by $N-M$ steps behind the
value of the master variable (if $N-M<0$, the slave map
anticipates the dynamics of the master map). To study the
stability of the synchronized solution we consider a perturbation
of the form
\begin{equation}
y_{n}= x_{n-N+M} + \delta_n .
\end{equation}

In the linear regime, the perturbation obeys the following map
\begin{equation}
\delta_{n+1}= f'(x_{n-N+M})\delta_n -\gamma \delta_{n-M},
\end{equation}
where $f'=df/dx$. Making the change of variables
$\delta_n=(z_n)^n$ gives
\begin{equation}
(z_{n+1})^{n+1}= f'(x_{n-N+M})(z _n)^n -\gamma (z_{n-M})^{n-M}.
\end{equation}

Notice that the value of $N$ is irrelevant in the long-term
behavior, thus the stability of the synchronized solution is the
same in the anticipated and in the retarded regimes (similar
results were found in [16]). Therefore, and without loss of
generality, in the following we consider $N=0$.

A sufficient condition for the stability of the synchronized
solution will be that, for $n>n_0$, where $n_0$ is some number of
transient steps, all the solutions $z_n^i$ of (8) (where $i$
labels the different solutions) satisfy $|z_n^i|<1$. When $f'$
depends on $x_n$, an analytical treatment of the stability of the
synchronized solution is in general not possible. However, there
are particular cases in which the study of Eq. (8) gives insight
into the parameter region where synchronization is stable. As a
first example we consider the 1D Bernoulli-like map
\begin{equation}
f(x_{n})=a x_n \ {\rm mod}\ 1 \ .
\end{equation}

The map is chaotic for $a>1$. $f'=a$, and thus the solutions of
(7) are linear combinations of functions of the form
$\delta_n=z^n$, with constant $z$. Thus (8) reads
\begin{equation}
z^{M+1}= a z^{M} -\gamma .
\end{equation}
Stability of the synchronized solution is obtained if all the
solutions of (9) satisfy $\vert z \vert <1$.

For $M=1$ the roots of Eq. (10) are the solutions of a simple
quadratic equation, so that it is simple to check that they have
$\vert z \vert <1$ when $\gamma \in (a-1, 1)$. For arbitrary $M$ a
{\it necessary} (but not sufficient) condition for the stability
of the synchronized solution is $\gamma \in [a-1, a+1]$. To show
that, first note that for $\gamma=0$, Eq. (10) has one root at
$z=a>1$, and the other $M$ roots are at $z=0$. A simple analysis
of perturbations shows that a small $\gamma$ breaks the degeneracy
of the roots at the origin, which then move radially outwards in
the complex plane, whereas (if $\gamma>0$) the root that was
located at $z=a$ diminishes its value, moving towards the unit
circle. Thus, stability of the synchronized solution will be
obtained by increasing $\gamma$ if this last root enters into the
unit circle in the complex plane before some of the other $M$
roots leave it. Finally, for $\gamma$ large enough, all the roots
are outside the unit circle. Eq. (10) can be rewritten as
\begin{equation}
z^M = -{{\gamma}\over{z - a} }.
\end{equation}

Thus, the value of $|z|^M$ is bounded between
\begin{equation}
{{\gamma}\over{a+|z|}} \le |z|^M \le {{\gamma}\over{a-|z|}}.
\end{equation}

At the limits of the stability region, some root $z$ would satisfy
$|z|=1$, so that
\begin{equation}
{{\gamma}\over{a+1}} \le 1 \le {{\gamma}\over{a-1}}.
\end{equation}
which leads to $a-1\le \gamma \le a+1$. Therefore, only within
this interval a root might cross the unit circle, leading to
stability changes of the synchronized solution. In consequence the
range of $\gamma$ leading to stable synchronization is inside
$\gamma \in [a-1, a+1]$. Further insight can be gained from the
study of the roots at the boundaries of this interval. First we
consider the case $\gamma=a-1$. For this value of the coupling Eq.
(10) becomes
\begin{equation}
z^{n+1}=az^n-a+1.
\end{equation}

Clearly, $z=1$ is a solution for all $a$ and $M$. A small
perturbation of the value of $\gamma$, $\gamma=a-1+\delta \gamma$
leads to a modification of the value of $z$, $z=1+\delta z$. To
first order, $\delta z$ and $\delta \gamma$ are related by
\begin{equation}
\delta z = - {{\delta \gamma} \over {1-M(a-1)}}.
\end{equation}

The denominator is positive if $M<1/(a-1)$ and negative otherwise.
Thus, if $ M<1/(a-1)$, by increasing $\gamma$, ($\delta
\gamma>0$), $\delta z <0$ and a real root enters into the unit
circle. Since this value of $\gamma$ is the smallest one for which
crossing the unit circle becomes possible, all the roots satisfy
now $|z|<1$ and thus the synchronized solution becomes stable. On
the other hand, if $M>1/(a-1)$, by increasing $\gamma$, $\delta
z>0$, and a real root leaves the unit circle. In this case the
synchronized solution becomes more unstable, in the sense that the
rate of escape given by the largest $|z|$ increases by increasing
$\gamma$. Thus, $M<1/(a-1)$ gives a limit of the number of steps
$M$ for which the synchronized solution can become stable. There
is a relation between $M$ and the degree of chaos, associated to
the Lyapunov exponent $\log a$, of the master map: the largest the
value of $a$, the lower the stable anticipation times. If $a>2$,
stable anticipated synchronization becomes impossible.

Next, we consider the other boundary of the synchronization
region, $\gamma=a+1$. In this case Eq. (10) becomes
\begin{equation}
z^{n+1}=az^n-a-1.
\end{equation}

If $M$ is even, $z=-1$ is a solution. Considering a perturbation
of the form $\gamma=a+1+\delta \gamma$, in the same way as before
it can be shown that $z=-1+\delta z$ with $\delta z = -\delta
\gamma/[1+M(a+1)]$. Since the denominator is always positive, if
$\gamma$ grows, $z$ decreases, so that the root close to $z=-1$
leaves the unit circle. Thus, we simply confirm that a necessary
condition for the stability of the synchronized solution is
$\gamma<a+1$.

Next we show results of numerical simulations that confirm these
analytic arguments. Figure 1 shows simulations of Bernoulli-like
maps, in which the slave anticipates the master in five steps.
During the first 5000 steps the maps evolve independently. Then we
set the value of the coupling to $\gamma=0.15$ and after a very
short transient the anticipation of the slave to the master is
evident. Figure 2 shows, for the same value of the parameters $a$
and $M$, and increasing coupling $\gamma$, how the roots of Eq.
(10) move in the complex plane. For low coupling Eq. (10) has one
real root $z_1
\sim a>1$, one real root $z_2 < 1$, and four complex conjugate
roots with modulus less than 1 [Fig. 2 (a)]. As the coupling
increases, $z_1$ decreases while the other roots increase their
modulus, approaching the unit circle. For $\gamma \gtrsim 0.1 =
a-1$ all roots of Eq. (8) have modulus less than 1 [Fig. 2 (b,c)]
and the synchronized solution is stable (Fig. 2(c) corresponds to
Fig. 1). For even larger coupling, pairs of complex-conjugate
roots cross the unit circle [Fig. 2 (d)], and synchronization is
unstable again.

For values of the coupling such that all roots of Eq. (10) are
inside the unit circle, the distance between the two trajectories
decreases exponentially, $|x_{n+M}-y_{n}| \sim |x_{M}-y_0| \exp
(-n/\tau)$, with the transient time to synchronization given by
the inverse of the logarithm of the modulus of the largest root,
$\tau=-1/\ln|z_1|$. Fig. 3 shows the transient evolution of
$|x_{n+M}-y_n|$, for the parameters of Fig. 1. We observe a damped
oscillatory behavior. While the damping time is $\tau$, the
frequency is associated to the phase of $z_1$.

\begin{figure}
\centering
\resizebox{.8\columnwidth}{!}{\includegraphics{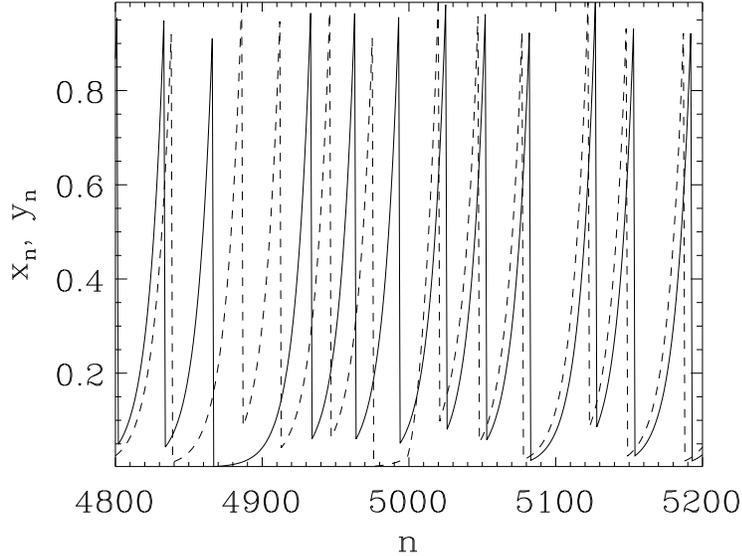}}
\caption{Time series of Bernoulli-like maps ($x_n$ solid line,
$y_n$ dashed line) for $a=1.1$, $\gamma=0.15$, $M=5$. The coupling
is set on at $n=5000$, leading to anticipated synchronization. }
\label{f1}
\end{figure}

\begin{figure}
\centering
\resizebox{.8\columnwidth}{!}{\includegraphics{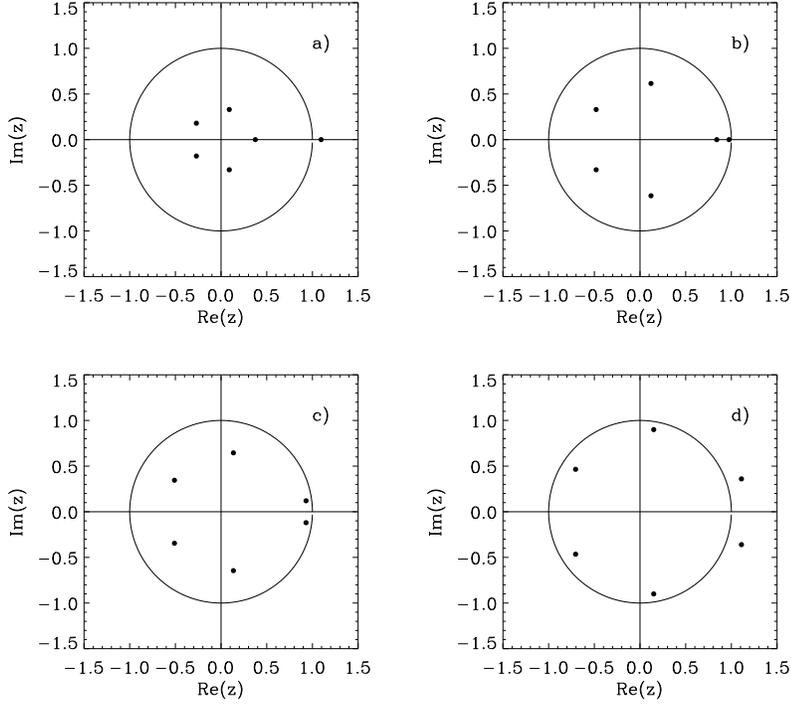}}
\caption{Roots of Eq. (10) for $a=1.1$, $M=5$ and (a)
$\gamma=0.005$, (b) $\gamma=0.11$, (c) $\gamma=0.15$,  and (d)
$\gamma=0.8$. }\vskip 0.5cm
\label{f2}
\end{figure}

\begin{figure}
\centering
\resizebox{.8\columnwidth}{!}{\includegraphics{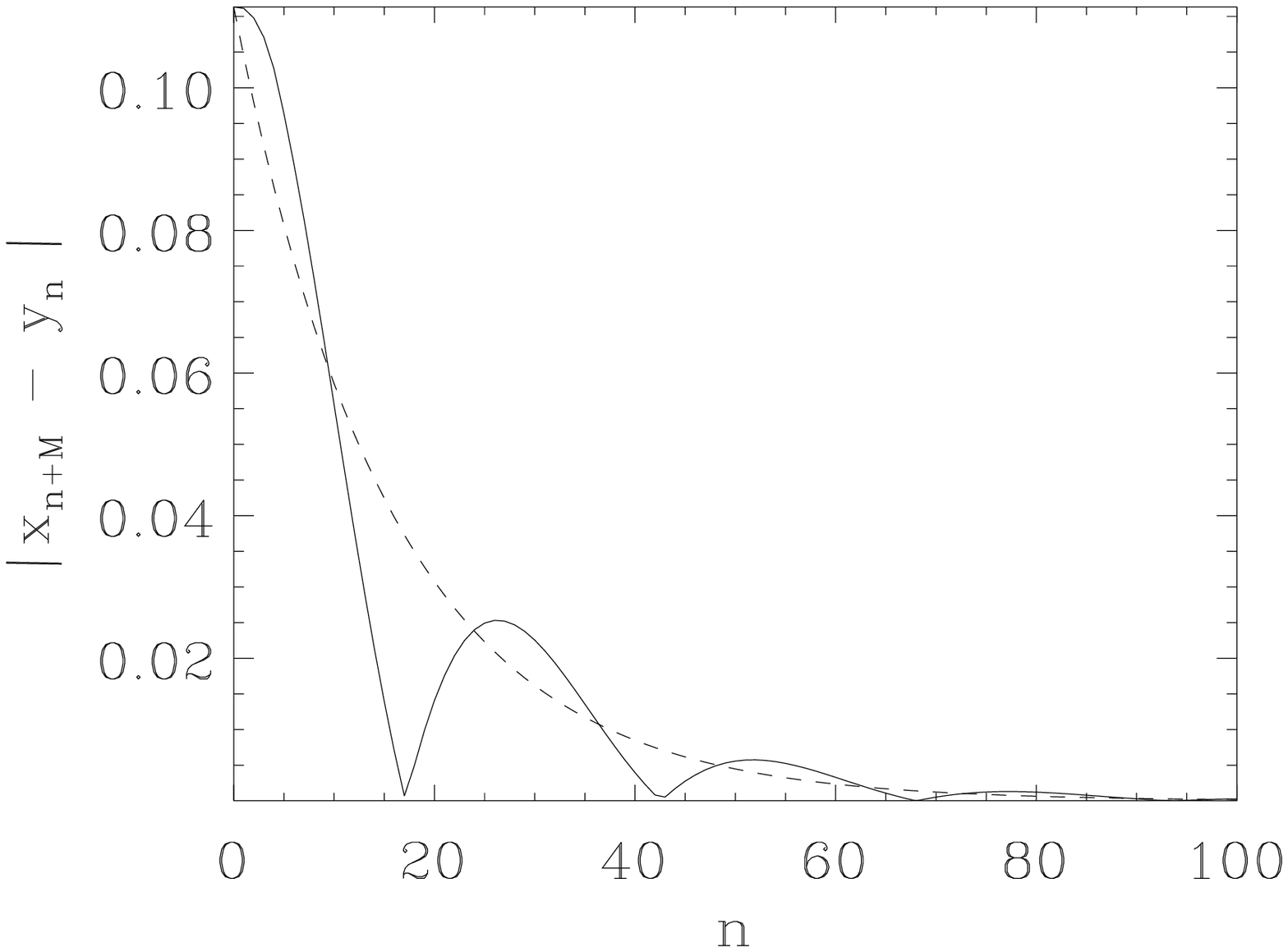}}
\caption{Transient decay of $|x_{n+M}-y_{n}|$ for $a=1.1$, $M=5$
and $\gamma=0.15$. The solid line indicates the value of
$|x_{n+M}-y_{n}|$ and the dashed line indicates the value of
$|x_{M}-y_0| \exp (-n/\tau)$ with $\tau=-1/\ln|z_1| \sim 15.55$ as
explained in the text. }\vskip 0.5cm \label{f3}
\end{figure}

As a second example we consider the 2D Baker's map that transform
the unit square into two non overlapping rectangles: The master
map, ($x_n^m,y_n^m$), and the slave map, ($x_n^s, y_n^s$), are
\begin{equation}
\left\{
\begin{array}{ll}
x^m_{n+1}= f_x(x^m_{n},y^m_{n})
\\ y^m_{n+1}= f_y(x^m_{n},y^m_{n}) \\
\end{array}
\right.
\end{equation}
\begin{equation}
\left\{
\begin{array}{ll}
x^s_{n+1}= f_x(x^s_{n},y^s_{n}) + \gamma(x^m_n - x^s_{n-M}) \\
y^s_{n+1}= f_y(x^s_{n},y^s_{n}) + \gamma(y^m_n - y^s_{n-M}).
\end{array}
\right.
\end{equation}
where
\begin{equation}
f_x = \left\{
\begin{array}{ll}
ax_n & \mbox{if $x_n<1/a$,}
\\ a(x_n-1/a) & \mbox{if $x_n \geq 1/a$,}
\\
\end{array}
\right.
\end{equation}

\begin{equation}
f_y = \left\{
\begin{array}{ll}
by_n & \mbox{if $x_n<1/a$,}
\\ by_n +(1-b) & \mbox{if $x_n \geq
1/a $,} \\
\end{array}
\right.
\end{equation}
$a>1$ and $b<1$ are the expansion and contraction rates,
respectively. It is easy to see that for the stability of the
synchronized solution we obtain a pair of equations of the same
form as Eq. (10), where now $a$ is equal to the expansion and to
the contraction rate, respectively. Both equations must have roots
with modulus less than 1, for the synchronized solution to be
stable. In Fig. 4 we present results of numerical simulations that
show anticipated synchronization by one step. The coupling is set
on at $n=30$, and after a transient the map solution approaches
the anticipated synchronization state. We find synchronized
solutions for parameter values such that the equations analogous
to (10) have roots with modulus less than 1. The duration of the
transient, again, is related to the modulus of the largest root.

\begin{figure}
\centering
\resizebox{.8\columnwidth}{!}{\includegraphics{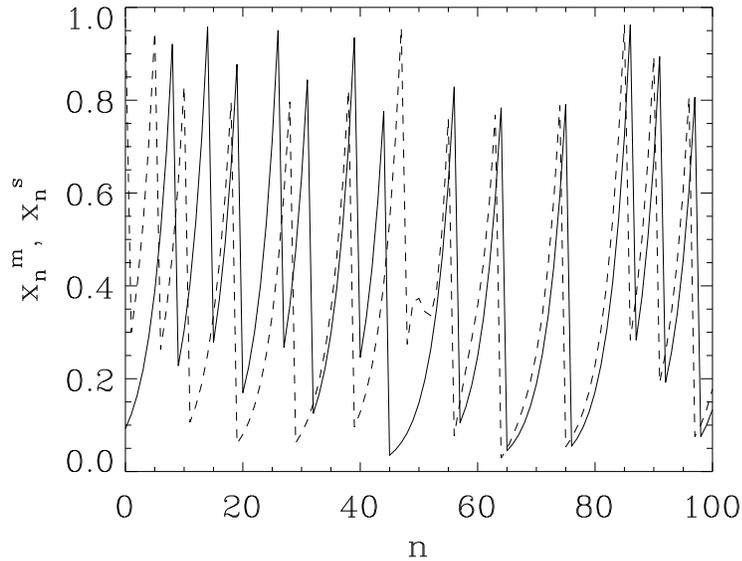}}
\caption{Time series of the 2D Baker map ($x^m_n$ solid line,
$x^s_n$ dashed line) for $a=1.333$, $b=0.777$, $\gamma=0.7$, and
$M=1$. The coupling is set on at $n=30$, leading to anticipated
synchronization. } \label{f4}
\end{figure}

In summary, we have studied the regime of anticipated
synchronization in unidirectionally coupled chaotic maps. In a
general case is not possible to give analytic conditions for the
parameter region where anticipated synchronization occurs, but
we have presented two class of maps, 1D Bernoulli-like maps and
2D Baker maps, in which an analytic treatment of the stability
of the synchronized solution is possible. The results of
numerical simulations are in good agreement with the analytic
predictions.

C. Masoller was supported by Proyecto de Desarrollo de Ciencias
B\'asicas (PE\-DE\-CI\-BA), Comisi\'on Sectorial de
Investigaci\'on Cient\'{\i}fica (CSIC), U\-ru\-guay, and
Universitat de les Illes Balears, Spain. E.
Hern\'{a}ndez-Garc\'{\i}a and C. Mirasso acknowledge support from
MCyT (Spain), project CONOCE BMF2000-1108.

\end{document}